\documentclass[pre,a4paper,oneside,showpacs,showkeys,twocolumn]{revtex4}

\usepackage{graphicx}
\usepackage{latexsym}   
\begin{document}

\title{Topological features of proteins from amino acid residue networks}
\author{\firstname{Nelson} A. \surname{Alves}}
\email{alves@ffclrp.usp.br} 
\author{\firstname{Alexandre} S. \surname{Martinez}}
\email{asmartinez@ffclrp.usp.br}
\affiliation{Departamento de F\'{\i}sica e Matem\'{a}tica, \\
             Faculdade de Filosofia, Ci\^encias e Letras de Ribeir\~ao Preto, \\
             Universidade de S\~ao Paulo \\ 
             Avenida Bandeirantes, 3900 \\ 
             14040-901, Ribeir\~ao Preto, SP, Brazil.}
\date{\today}

\begin{abstract}
Topological properties of native folds are obtained from statistical analysis of 160 low homology proteins covering the four structural classes.  
This is done analysing one, two and three-vertex joint distribution of quantities related to the corresponding network of amino acid residues. 
Emphasis on the amino acid residue hydrophobicity leads to the definition of their center of mass as vertices in this contact network model with interactions represented by edges. 
The network analysis helps us to interpret experimental results such as hydrophobic scales and fraction of buried accessible surface area in terms of the network connectivity. 
To explore the vertex type dependent correlations, we build a network of hydrophobic and polar vertices.
This procedure presents the wiring diagram of the topological structure of globular proteins leading to the following attachment  probabilities between hydrophobic-hydrophobic 0.424(5), hydrophobic-polar 0.419(2) and polar-polar 0.157(3) residues. 
\end{abstract}


\maketitle

\section{Introduction}

It has long been recognized that limitations on modeling proteins are due to the complexity of the intra-molecular interactions and the lack of an adequate description of the interactions with the solvent~\cite{FeigCOSB04}.
The complexity of the interactions has led to sophisticated numerical simulations of all-atom models. 
Thus, it is primordial to consider the basic features of the protein dynamics to obtain a reliable and less computationally demanding working model to understand the folding process and protein stability.  
Many theoretical models have been proposed to the protein structure prediction with reduced degrees of freedom~\cite{KolinskiPolymer04}.
This leads to a coarse-grained approach where all-atom interactions are replaced by centers of force.

Beyond the dynamical approach, many natural and artificial interacting systems have been modeled as a graph, where vertices represent the system elements and edges represent their interactions. 
In particular, some graphs present scale free behavior and are called complex networks~\cite{BarabasiRev,DorogovtsevRev,newman_2003}.
Theoretical studies have focused on the understanding of their large-scale organization~\cite{Strogatz98,amaral_2000} and growing mechanism~\cite{Barabasi99}, possibly driven by evolutionary principles~\cite{Gopal01,valverde_2002,colizza_2004}.

In this view, networks of interacting amino acid pairs have been constructed given the Cartesian coordinates of the main chain $\alpha$-carbon atoms as vertices.
Edges are established when any other $C^\alpha$ atom is within a cutoff Euclidean distance $R_c$~\cite{Vendruscolo,greene_2003,Atilgan,bagler_2005,Kundu}.
Distances as $R_c=8.5$ \AA~\cite{Vendruscolo,greene_2003,Dokholyan02} or $R_c=7.0$ \AA~\cite{Atilgan,bagler_2005}, have been taken to define a contact network of amino acids.

Another approach to model an amino acid network can be devised if one considers the center of mass of their side chains as vertices.
Those centroids are obtained with the heavy atoms spatial coordinates of the amino acid residues and act as interaction sites.
Again, a contact network is defined if a centroid is within an Euclidean distance $R_c=8.5$ \AA\ of another one and this case is studied in this paper. 
This approach emphasizes the amino acid residue hydrophobicity, which is the main force leading to a folded conformation~\cite{Dill90,southall_2002}.
 
With this network modeling, following the work by Greene and Higman~\cite{greene_2003}, we are able to gather quantitative results characterizing the native conformation topology of globular proteins.  
As we show, this procedure helps us to rationalize the formation of native states from a topological point of view. 
This is achieved with a network description for the inter-residue interactions, which helps to interpret experimental results such as hydrophobic scales and the fraction of buried accessible surface area. 
Moreover, we evaluate the attachment probabilities between hydrophobic-hydrophobic, hydrophobic-polar and polar-polar residues from this simple network model. 

The definitions of the topological quantities used in this paper are presented in Sec.~\ref{results_discussion} with the numerical results followed by a discussion. 
The final conclusions are presented in Sec.~\ref{conclusion}. 
Appendix~\ref{apendice} complements our study through the distribution of inter-residue  contacts.  

\section{Results and Discussion}
\label{results_discussion}

To explore the network model of globular proteins, statistical analysis have been carried out for networks built from 160 high-resolution and low homology structures, presenting different folds and chain lengths~\cite{dataset}, collected in the Protein Data Bank (PDB). 
Chains with residues omitted in the PDB database have not been included in our data set. 
This avoids bias in the calculation of topological quantities.
Our data consist of four sets of 40 different native-state conformations representing the four broad structural classes, all-$\alpha$, all-$\beta$, $\alpha + \beta$ and $\alpha / \beta$ according to the SCOP classification~\cite{SCOP}.
The selection of protein structures has been made to permit a numerical analysis of the dependence of network topological properties on secondary structure contents and their network size $N$, i.e., the total number of amino acids. 

\begin{figure}[t]
 \begin{center}
 \includegraphics[angle=-90,width=\columnwidth]{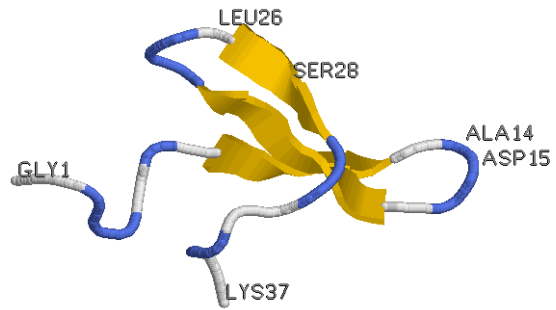}
 {\bf(a)}
 \includegraphics[angle=0,width=\columnwidth]{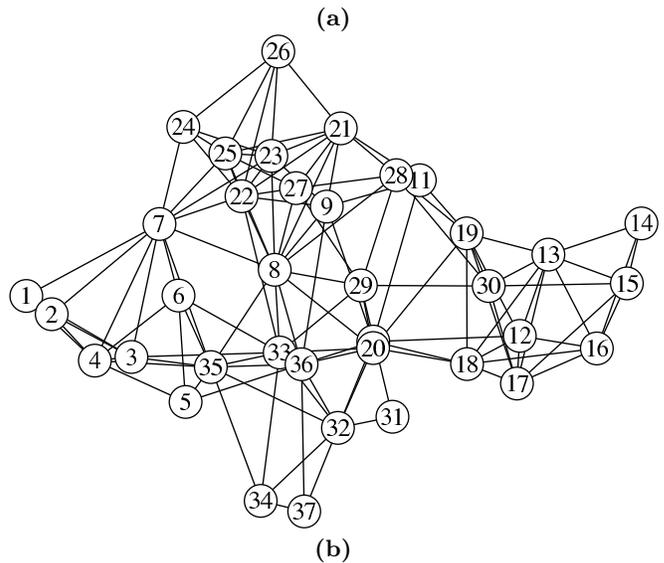}
 {\bf(b)}
 \end{center}
\caption{Example of the amino acid residue network for the PDB code 1E0L, a small 37-amino acid protein in the all-$\beta$ class. 
Figure {\bf (a)} shows the ribbon diagram obtained with RASMOL visualization tool.
Figure {\bf (b)} displays the corresponding network with their numbered vertices representing the center of mass of the side chains. 
This figure was drawn using VISONE (www.visone.de).
}
\label{fig:1}
\end{figure}

Figure~\ref{fig:1} shows an illustrative example of the network structure with the vertices calculated from side chain centroids for the protein 1E0L, collected in our all-$\beta$ data set. 

\subsection{One vertex analysis: mean network degree and hydrophobicity scale}
\label{sec:one_vertex}

\begin{figure}[t]
 \begin{center}
 \includegraphics[angle=0,width=\columnwidth]{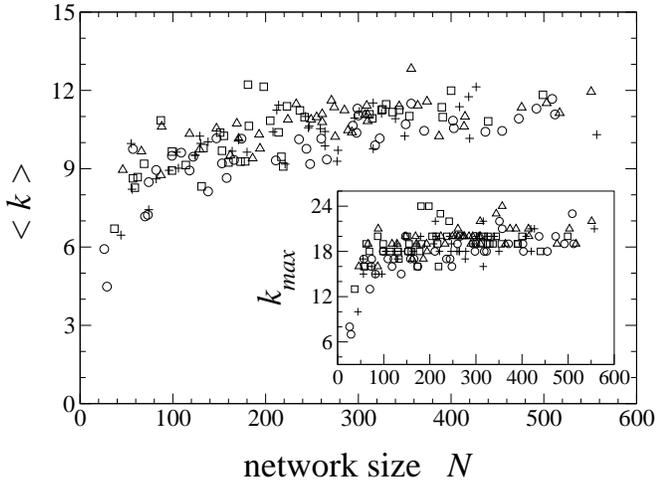}
 \end{center}
\caption{Average connectivity as a function of the network size $N$ for the structural classes: all-$\alpha$ ($\circ$), all-$\beta$ ($\Box$), $\alpha + \beta$ (+) and $\alpha / \beta$ ($\triangle$). 
  {\bf Inset:} the maximum connectivity $k_{max}$ for each network as a function of $N$.}
\label{fig:2}
\end{figure}

Figure~\ref{fig:2} shows the average vertex (degree) connectivity  $\langle k \rangle$ for the four structural classes as a function of $N$.
It is obtained as an average over the connectivities $k_i$, where $k_i$ is the number of edges connecting the vertex $i$ to other vertices in the network. 
The value of $\langle k \rangle$ increases quickly for small proteins and shows an asymptotic trend for all classes.
This figure indicates a predominant lower average connectivity for all-$\alpha$ conformations compared with all other classes.

The typical $\langle k \rangle$ for proteins are to be compared with other biological
systems~\cite{BarabasiRev}:
{\it S. cerevisiae} protein-protein network,  $\langle k \rangle = 2.39$  and $N=1870$;  
{\it E. coli} metabolic network,  $\langle k \rangle = 7.4$  and $N=778$;  
neural network of {\it C. elegans},  $\langle k \rangle = 14$ and $N=282$; 
Silwood Park food web, $\langle k \rangle=4.75$  and $N=154$;  
Ythan estuary food web, $\langle k \rangle= 8.7$  and $N=134$.

A protein mapping onto a contact network clearly introduces constraints on the network topological properties.
The first one is due to the finite size of real data, which introduces a cutoff degree depending on $N$~\cite{Pastor04}.
Second, size and steric considerations introduce a more relevant constraint in the topology of amino acid networks, for example, limiting the possible values for the degree $k$.
Thus, the degree dependence must be characterized by a limiting finite value for the maximum degree $k_{max}$, irrespective of the chain size.  
This can be observed for our data sets in the inset of Figure~\ref{fig:2}.
The hitting of the maximum degree of a vertex can be translated as performance of how to connect vertices in the folding process.
Once this value is reached, the network rewires edges connected to that vertex or stops this specific vertex dependent process.

This conclusion is also reinforced by a similar study carried out by Vendruscolo \emph{et al.}~\cite{Vendruscolo} with the ``betweeness'' $B_i$, which correlates with $k_i^2$, when comparing native and Monte Carlo generated transition states. 
As explicitly demonstrated for six proteins, highly connected residues in the native state may not correspond to the highly connected ones (key residues) in the transition states. 

Protein structure is sequence dependent. 
Thus, different of other real network models, where vertices are in general considered to be of
the same type we are dealing with systems, which result from the intrinsic properties of the amino acid residues.

\begin{figure}[t]
 \begin{center}
 \includegraphics[angle=0,width=\columnwidth]{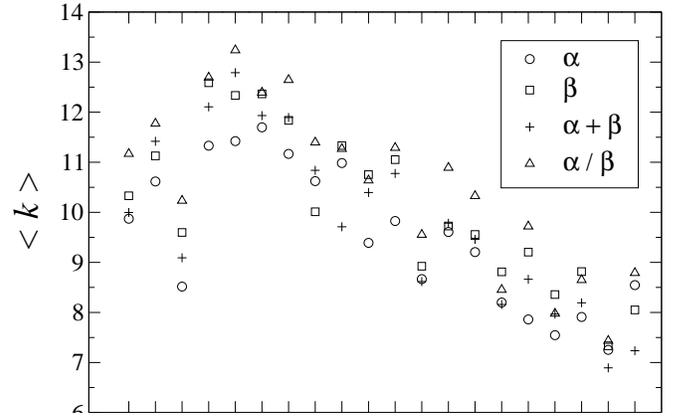}
 {\bf (a)} 
 \vspace{1cm}
 \includegraphics[angle=0,width=\columnwidth]{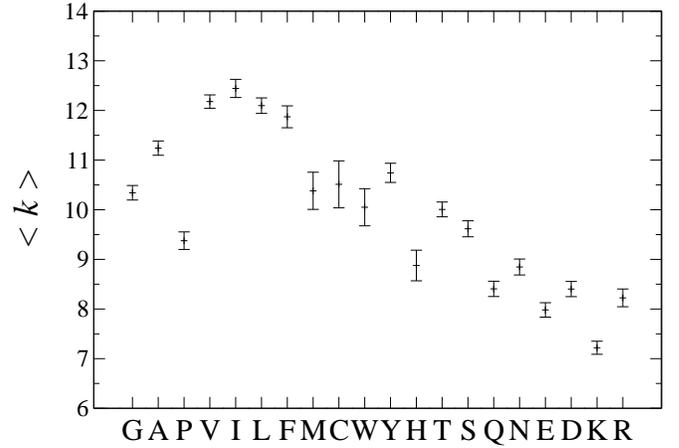}
 {\bf (b)}
 \end{center}
\caption{{\bf (a)} Average connectivity of each amino acid residue according to the structural class.
         {\bf (b)} Average connectivity and standard deviation of each amino acid residue calculated from all networks.
         The four more connected residues are Val (V), Ile (I), Leu (L) and Phe (F), while the less connected ones are Gln (Q), Glu (E), Asp (D), Lys (K) and Arg (R).}
\label{fig:3}
\end{figure}

To obtain a deeper insight of the role played by each residue in forming
native conformations, we investigate how to relate their character
to the ability of making contacts with other residues. 
For this end, we display in Fig.~\ref{fig:3} the average connectivity for each residue one-letter code from all 160 structures, taking into account the number of times each residue appears in all sequences.

The sequence of amino acid residues displayed in Figure~\ref{fig:3} was chosen because it groups their characteristics: aliphatic non-hydrogen-bonding (Gly, Ala, Pro, Val, Ile, Leu), aromatic non-hydrogen-bonding (Phe), sulfur containing (Met, Cys) and hydrogen-bonding (Trp, Tyr, His, Thr, Ser, Gln, Asn, Glu, Asp, Lys, Arg)~\cite{PKarplus}.

Figure~\ref{fig:3}a shows a clear trend for lower average connectivity played by amino acid residues in proteins classified in all-$\alpha$ structural class. 
Nevertheless, the highest average connectivity is presented by residues in the $\alpha/\beta$ protein class. 
This class has mixed $\alpha$-helices and parallel $\beta$-strands. 
Intermediary pattern of connectivity is found for residues in all-$\beta$ and $\alpha + \beta$ protein classes with predominant higher connectivity for the $\beta$ class. 
Therefore, it is likely that the higher average connectivity for the residues in $\alpha/\beta$ class also comes from interactions among residues in different secondary structures. 
The $\alpha + \beta$ class presents regions of $\alpha$-helices and anti-parallel $\beta$-strands largely segregated along the chain. 
We also conclude that the connectivity in this protein class results mainly from contacts among the elements in each secondary structure type. 
Figure~\ref{fig:3}b shows the average connectivity of each residue calculated from all networks considered in our data set, stressing their general ability of making network contacts.

Can we relate this ability to any intrinsic amino acid property? 
In the following, we argue that those sharp evaluated average values
can be understood in terms of the concept of hydrophobicity.

We recall that the hydrophobic effect concept as the leading force in folding globular proteins comes from experimental evidences.
This is illustrated by the finding of mostly non-polar residues in the core of protein structures~\cite{Dill90,southall_2002} as a consequence of avoiding the exposure to water of amino acids with hydrophobic side chains.
Therefore, the hydrophobic/hydrophilic character of amino acids has been viewed as an important issue in understanding the formation of native structures.
For example, the spatial distribution of hydrophobic and hydrophilic residues has led to the definition of score functions based on hydrophobic profiles as a mean to detect native protein folding among many decoy structures~\cite{Silverman03,alves_2005}.

In spite of many scales ranking the hydrophobic character, due mainly to the particular measurement method, there is a consensus about the more hydrophobic (Phe, Ile, Leu, Val) and more hydrophilic (Lys, Arg, Asp, Glu) residues.
Therefore, one expects to obtain more (less) connected vertices as the ones ranked in the top (bottom) of the hydrophobic scales.

To investigate how the average network connectivity (shown in Figure~\ref{fig:3}b) is related to the hydrophobic character of residues, we have checked its correlation with the 127 scales collected in Ref.~\cite{Palliser_scales}.
The most significant correlations are observed with the well known scales in the literature: Kyte and Doolittle, Juretic \emph{et al.} and Janin-2, all presenting correlation coefficient $\rho=0.94$; Taylor and Thornton ($\rho= 0.93$) and Sereda \emph{et al.}-2 ($\rho=0.92$), following the notation for these scales as in Table I of Ref.~\cite{Palliser_scales}.
On the other hand, non-significant correlations are obtained with the scales: Hopp ($\rho=0.10$); Wilce \emph{et al.}-3 ($\rho= 0.14$);  Michel \emph{et al.}, Colonna-Cesari and Sander ($\rho = 0.21$). 

Although we can find a very good agreement between the amino acid hydrophobic caracter and its average connectivity, Figure~\ref{fig:3}(a) clearly shows that this quantitative character is related to the secondary structure contents. 
Yet, this has experimental consequences on the definition of the free energy transfer. 
Therefore, any method used to predict elements of secondary structure based on typical hydrophobicity profiles need to be reconsidered. 
Those quantitative values are expected to be correlated to the exposure pattern of the amino acids to the aqueous environment. 
This can be readly investigated considering only the average connectivity and the available data on  individual amino acid accesible surface area (independent of its location in a peptide).

Our results show that the average connectivity reveals a weak correlation with the average buried accessible surface area (ASA) $A^0 - \langle A \rangle$ ($\rho= 0.39$)~\cite{HZhouYZhou,HZhouYZhou2}, where $A^0$ is ASA calculation based on Gly-X-Gly models related to the unfolded state and $\langle A \rangle$ is an average ASA calculated from native protein structures. 
However, the correlation coefficient increases to 0.90 for the \emph{fraction} of buried ASA $1 - \langle A \rangle/A^0 $, which reflects the relative loss of their accessible surface areas during the folding process.

\subsection{Two vertices analysis: assortativity}

So far we have focused the discussion from a single vertex point of viewing.
However, many real-world networks also exhibit degree correlations~\cite{NewmanPRL02,PastorPRL01}, i.e., the conditional probability $P(k'|k)$ that an edge leaving from a vertex, with degree $k$, and arriving at a vertex, with degree $k'$, is dependent on $k$.

This two-vertices correlation is expressed by $\overline{k}_{nn}(k) = \sum_{k'} k'\,P(k'|k)$ and furnishes the average connectivity of the nearest neighbors of vertices with connectivity $k$~\cite{PastorPRL01,PastorPRE03,vasquez_2003}.
Concerning this aspect, the networks are classified as to show assortative mixing, if the degree correlation is positive, a preference for high-degree vertices to attach to other high-degree vertices, or disassortative mixing, otherwise.  
Therefore, whenever $\overline{k}_{nn}(k)$ increases with $k$, one obtains a correlated network, thus showing an assortative mixing. 
On the other hand, a decreasing behavior as a function of $k$ shows a disassortative mixing. 
 
Our numerical analysis of the two-vertices correlation shows that the dependency of $\overline{k}_{nn}(k)$ on the degree $k$ can be described by a linear relation for all our networks (data not shown), $\overline{k}_{nn}(k)= a + b\,k$,  with $a,b > 0$.
Moreover, this result leads to the average of $\overline{k}_{nn}(k)$ over all vertices, $ \langle \overline{k}_{nn}\rangle_N = \sum_{k} P(k) \overline{k}_{nn}(k)$,
where $P(k)$ is the observed degree distribution of each network, to have also a linear behavior as a function of  $\langle k \rangle$ because $\langle k \rangle = \sum_k k P(k)$.

\begin{figure}[ht]
 \begin{center}
 \includegraphics[angle=0,width=\columnwidth]{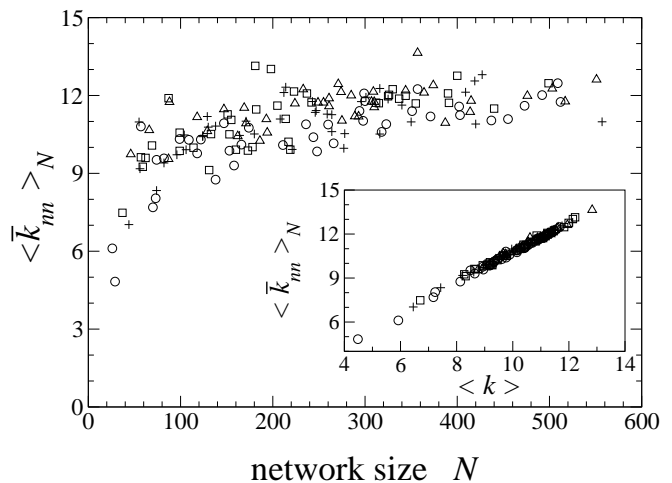}
 \end{center}
\caption{Average of $\overline{k}_{nn}(k)$ over all vertices of degree $k$, $\langle \overline{k}_{nn}\rangle_N$, as a function of the network size $N$ for the structural classes:  all-$\alpha$ ($\circ$), all-$\beta$ ($\Box$),  $\alpha + \beta$ (+) and $\alpha / \beta$ ($\triangle$). 
         {\bf Inset:} $\langle \overline{k}_{nn}\rangle_N$ as a function of the average connectivity $\langle k \rangle$ for all structural classes.
  }
\label{fig:4}
\end{figure}

We depict in Figure~\ref{fig:4} the behavior of  $\langle \overline{k}_{nn}\rangle_N$ as a function of $N$ and of the structural classes.
We observe again a stationary value for large $N$, reflecting what we can call the main property of amino acid residue network. 
In fact, that homogeneous network behavior ($\langle \overline{k}_{nn}\rangle_N \sim \langle k \rangle  $) is supported by the inset of Figure~\ref{fig:4}.
The linear regression for the 160 proteins leads to $\langle \overline{k}_{nn}\rangle_N = 0.65(7)+ 1.012(7) \langle k \rangle$, with correlation coefficient $\rho = 0.99$.  
Therefore, there is a strict correspondence, exhibited by the above equation, between the way $\langle \overline{k}_{nn}\rangle_N$ and the average degree $\langle k \rangle$ grow with $N$.
As a matter of fact, this can also be noted \emph{a posteriori} by the similarity between the positions of the data points in Figures~\ref{fig:2} and~\ref{fig:4}.  
This reveals a specific pattern of attachments among vertices for all network sizes, and it is further explored below.   

We recall that, although we observe a linear increasing of $\overline{k}_{nn}(k)$ with  $k$, this behavior is only valid up to $ k \sim  k_{max} $ due to physical constraints of the networks.

\subsection{Hydrophobic and hydrophylic vertices and their correlation}

So far, we have reinforced the observed process toward some high-connected vertices as the final result of the protein folding within its biological time.
However protein folding corresponds to a specific mechanism of wiring in contrast to the collapse of homopolymers, likely due to the key residues in according to the nucleation-condensation model as a highly cooperative process~\cite{Vendruscolo,fersht_pnas}. 

To explore further the role played by the connectivity of residues leading to the assortative mixing behavior we re-define our networks as being composed of two types of residues, denoted by $H$ and $P$.
$H$ residues are chosen as the more connected ones in Figure~\ref{fig:3}b and $P$ residues otherwise, in such way to obtain two data sets of 11 and 9 residues, roughly corresponding to the three more correlated scales we have identified in Sec.~\ref{sec:one_vertex}.
This makes a set for $H$ type residues including Thr as being the lowest hydrophobic and another one for $P$ type residues starting with Ser. 
Differently from bipartite networks, where edges connect only vertices of unlike type, our new networks have edges connecting vertices irrespective of their type.
Now, to proceed the investigation on how the assortativity arises in native structures we apply the ideas of community forming by Newman~\cite{NewmanPRE6703}. 
In this sense, we search a better comprehension of the assortativity as consequence of a mechanism for network formation,  which results from preferential attachments among specific residues.
Firstly, we quantify the observed assortative mixing calculating the fraction of edges $e_{ij}$ between vertices of type $i$ and $j$ ($i, j = \{H,P\}$). 
Table I presents the averaged values with their standard deviations for the fraction of edges in each structural class and the final averages from all data, with  $e_{HP} = e_{PH}$. 
It is clear the preference for attachments between $H-H$ and $H-P$ vertices.
Secondly, we calculate the assortativity coefficient~\cite{NewmanPRL02,NewmanPRE6703} $r$, a correlation coefficient which lies in the range $[0, 1]$.
This coefficient hits the value 1 for perfect assortative mixing, i.e. when every edge connects vertices of the same type, and the minimum value 0 for a random mixing.
The low numerical values for $r$ in Table I taking into account their standard deviations confirms that there is almost no assortative mixing by type, i.e. $H-H$ and $P-P$ attachments.
Hence, the network surrounding residue of type $H$ is likely to be another $H$ or $P$, with probabilities presented in Table I.
We may interpret those sharp probabilities as the result of a compositional equilibrium of $H$ and $P$ residues in globular
proteins. 
With our hydrophobic scale defined in terms of $\langle k \rangle$ we calculate how dominant the presence of $H$ residues is.
For all-$\alpha$ class we obtain $0.549 \pm 0.007$; all-$\beta$, $0.569 \pm 0.008$; $\alpha+\beta$, $0.559 \pm 0.007$ and $\alpha/\beta$, $0.579 \pm 0.006$. 

Although the tendency of $H$ residues to connect among themselves and with $P$ residues is a well known fact, this approach presents a manner to quantify the preferential attachment associated to globular proteins.
Furthermore, the tuning of energy couplings in minimal model simulations may profit from this residue attachment quantification in order to be more realistic in describing the folding process.    

\begin{table*}[ht]
 \begin{center} 
 \begin{tabular}{ccccc}
 \hline
                    &  $e_{HH}$  &  $e_{HP}$  & $e_{PP}$    &  $r$     \\   
                   &            &            &             &          \\   
 \hline
 all-$\alpha$       &  0.405(9)  & 0.422(5)   & 0.173(7)    & 0.099(8) \\
 all-$\beta$        &  0.420(10) & 0.426(5)   & 0.154(6)    & 0.073(8) \\
 $\alpha +\beta$    &  0.428(9)  & 0.417(5)   & 0.155(5)    & 0.091(6) \\
 $\alpha / \beta$   &  0.443(7)  & 0.412(4)   & 0.145(4)    & 0.091(6) \\
 all classes and all sizes      &0.424(5)    &0.419(2)   & 0.157(3)  & 0.089(4) \\
 all classes with $N < 100$     &0.419(18)   &0.424(9)   & 0.157(11) & 0.070(13) \\
 all classes with $100 \leq N < 200$ &0.438(9) &0.410(5) & 0.152(7)  & 0.096(8) \\
 all classes with $N \geq 200$ &0.420(5)   &0.422(2)   & 0.159(3)  & 0.091(4) \\
 \hline
 \end{tabular}
 \end{center}
\caption{Fraction of edges $e_{ij}$ between a vertex of type $i$ ($H$ or $P$) and a vertex of 
type $j$ ($H$ or $P$). Results are presented for globular proteins according to the structural classes and for all classes according to the number of amino acids $N$. 
The quantity $r$ is the assortativity coefficient.}
\end{table*}

From a graph point of view, attachment probabilities reveal the wiring diagram when including different kinds of vertices as a mechanism for network formation.
Those probabilities also discard the simple picture of a predominant hydrophobic core.
Our results help to form the image that an $H$ vertex brings along other $H$ vertices as well as other $P$ vertices in forming the native structure. 

Table I also shows the fraction of edges $e_{ij}$ for all classes as a function of three typical ranges of protein sizes $N$.
It is interesting to note that when one restricts the analysis for chain sizes in those ranges, the fractions $e_{ij}$ in each range are still comparable within the standard deviations.
This confirms the robust pattern of network formation irrespective of their size.

\subsection{Three vertices analysis: non-random pattern}

\begin{figure}[ht]
 \begin{center}
 \includegraphics[angle=0,width=\columnwidth]{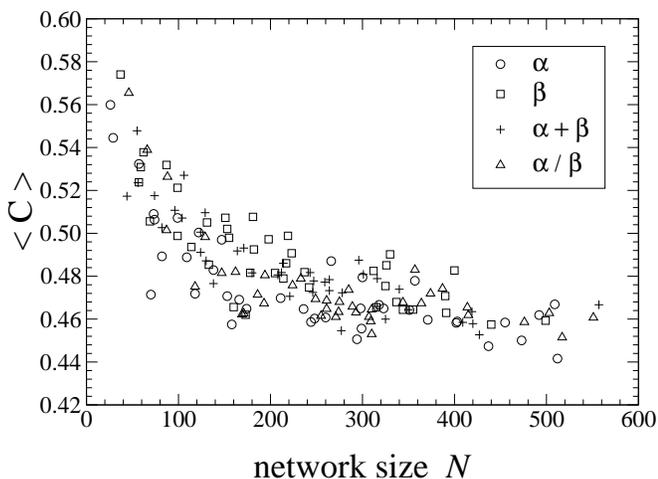}
 \end{center}
\caption{Clustering coefficient calculated with our network model as a function of $N$.}
\label{fig:5}
\end{figure}

To support the above specific attachment among $\{H,P\}$ residues, we have evaluated the clustering coefficient $\langle C \rangle$ over all vertices.
While the average connectivity gives information on how connected a vertex is, the clustering coefficient shows the network transitivity, i.e., the probability that two nearest neighbors of a given vertex are nearest neighbors themselves~\cite{Strogatz98}, forming a triangle with this vertex.
Notice that, in this aspect, the clustering coefficient is a measure of three-vertices correlation. 
The maximum value of $\langle C \rangle$ is 1 in a fully connected network, it assumes the value $\langle k \rangle/N$ for a random network and it is constant for regular lattices~\cite{NewmanStatP10100}. 
Our results for $\langle C \rangle$ as a function of $N$ for the structural classes exclude any random character in its formation (See Figure~\ref{fig:5}) as already pointed out~\cite{Vendruscolo,greene_2003} without including any size dependence analysis. 

Next, we look at the distribution of inter-residue contacts from a network perspective. 
However, since the main results of this distribution are already known in the literature, we have left them to the appendix. 
We include this appendix to show that network modelling distinguishes molecular chains with secondary structures from polymers and to present a comprehensive study of the topological properties of globular proteins.

\section{Conclusion}
\label{conclusion}

Different from the usual network analysis dedicated to study the properties of a single system, we try to infer the main topological features of a class of systems (globular proteins) characterized by an unknown mechanism which drive them to specific tridimensional configurations, the native structures. 

Topological parameters have been obtained to characterize native structures and their organization pattern as a result of the underlying dynamical process of formation. 
Our main results permit the following theoretical conclusions: (i) the average number of edges leaving a given center of force presents a direct relation with the hydrophobic character of the residues and to their fraction of buried accessible surface area; (ii) the way this average number of edges (center of forces) grows with the protein chain size is also reproduced in the formation of inter-residue interactions among the nearest neighbors of that center of force; (iii) the interactions among specific residues exhibit a well defined pattern described by the results presented in Table I.
The above patterns remain, irrespective of the protein size, and this may shed light on the protein evolution.

\acknowledgments

N. Alves thanks valuable discussions with U.H.E. Hansmann. 
The authors also acknowledge support from the Brazilian agencies CNPq (303446/2002-1 and 305527/2004-5) and FAPESP (2005/04067-6 and 2005/02408-0). 

\appendix

\section{Distribution of Inter-Residue Contacts}
\label{apendice}

\begin{figure}[ht]
 \begin{center}
 \includegraphics[angle=0,width=\columnwidth]{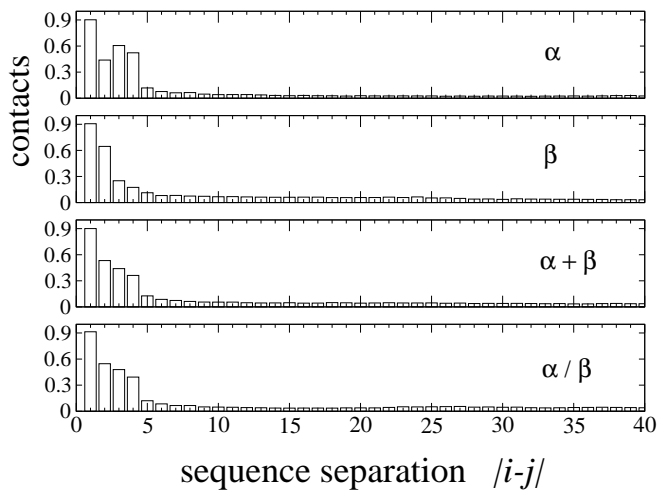}
 \end{center}
\caption{Average relative number of inter-residue contacts as a function of the sequence separation $|i-j|$.  Results are obtained for the four structural classes of proteins with $N \geq 200$.}
\label{fig:6}
\end{figure}

There is still another important aspect to be considered: the attachments distribution in our networks compared to a full connected one, as driven by bonded and non-bonded forces from inter-residue interactions. 
This can be translated as a topological pattern of inter-residue contacts categorized into short-, medium- and long-range interactions.
Short-range interaction contributions correspond to residues within a distance of two residues from each other, medium range to the ones within a distance of 3 or 4 residues and long-range contribution is defined to come from more than 4 residues away~\cite{Gromiha_Prog04}. 
Figure~\ref{fig:6} presents the average relative number of observed inter-residue contacts as a function of its sequence separation $|i-j|$ between residues $i$ and $j$ along the main chain.
It is defined as the ratio of the observed number of contacts at distance $|i-j|$ and its maximum number for the full connected network, i.e. with $N(N-1)/2$ edges.
This figure shows that the main contribution comes from the first four neighbors, i.e., from short- and medium-range contacts, reflecting the chain connectivity property. 
In particular, for all-$\alpha$ chains, we see that from all possible contacts at distance 1, only 90.2\% occur; at distance 2, 43.9\%; at distance 3, 60.6\%; at distance 4, 52.2\%; dropping to 11.8\% and 7.7\% at distances 5 and 6, respectively. 
Comparing all classes, the short-range contacts present similar relative numbers.
Large relative number (but less than 1) of contacts at distance 1 means that the chain connectivity between successive residues $(i, i+1)$ is not always enforced due to our definition of cutoff distance $R_c$ among side chain center of mass.  

We observe a preference for medium-range contacts by all-$\alpha$ structures and a very low preference by all-$\beta$.
The highest relative number of contacts for all-$\alpha$ at distance 3 reveals the pattern of helices characterized by the $(i,i+3)$ contacts.
Classes $\alpha+\beta$ and $\alpha/\beta$ present similar average relative number of observed inter-residue contacts.
It is clear the preference of long-range contacts for all-$\beta$ structures, mainly compared with all-$\alpha$ structures.
This overall behavior reveals how vertices related to secondary structures interact in a network of amino acids as a function of the sequence separation.

\clearpage


\begin{thebibliography}{10}

\bibitem{FeigCOSB04}
M.~Feig, C.~L. {Brooks III},  Curr. Opin. Struct.
  Biol. {\bf 14}, 217 (2004).

\bibitem{KolinskiPolymer04}
A.~Kolinski, J.~Skolnick, Polymer {\bf 45}, 511 (2004).

\bibitem{BarabasiRev}
R.~Albert, A.-L. Barab\'asi,  Rev.
  Mod. Phys. {\bf 74}, 47 (2002).

\bibitem{DorogovtsevRev}
S.~N. Dorogovtsev, J.~F.~F. Mendes,  Adv. Phys. {\bf 51}, 1079 (2002).

\bibitem{newman_2003}
M.~E.~J. Newman, SIAM Rev.
  {\bf 45}, 167 (2003).

\bibitem{Strogatz98}
D.~J. Watts, S.~H. Strogatz, 
  Nature {\bf 393}, 440 (1998).

\bibitem{amaral_2000}
L.~A.~N. Amaral, A.~Scala, M.~Barth\'el\'emy, H.~E. Stanley, Proc. Natl. Acad. Sci. U.S.A. {\bf 97},   11149 (2000).

\bibitem{Barabasi99}
A.-L. Barab\'asi, R.~Albert, Science 
  {\bf 286}, 509 (1999).

\bibitem{Gopal01}
N.~Mathias, V.~Gopal, Phys. Rev. E {\bf 63}, 021117 (2001).

\bibitem{valverde_2002}
S.~Valverde, R.~Ferrer-Cancho, R.~V. Sol\'e,  Europhys. Lett. {\bf 60}, 512 (2002).

\bibitem{colizza_2004}
V.~Colizza, J.~R. Banavar, A.~Maritan, A.~Rinaldo,  Phys. Rev. Lett. {\bf 92}, 198701 (2004).

\bibitem{Vendruscolo}
M.~Vendruscolo, N.~V. Dokholyan, E.~Paci, M.~Karplus,  Phys. Rev. E {\bf 65}, 061910 (2002).

\bibitem{greene_2003}
L.~H. Greene, V.~A. Higman,  J. Mol. Biol. {\bf 334}, 781 (2003).

\bibitem{Atilgan}
A.~R. Atilgan, P.~Akan, C.~Baysal, Biophys. J. {\bf 86}, 85 (2004).

\bibitem{bagler_2005}
G.~Bagler, S.~Sinha, Physica A {\bf 346}, 27 (2005).

\bibitem{Kundu}
S.~Kundu, Physica A {\bf 346}, 104 (2005).

\bibitem{Dokholyan02}
N.~V. Dokholyan, L.~Li, F.~Ding, E.~I. Shakhnovich, Proc. Natl. Acad. Sci. U.S.A. {\bf 99},  8637 (2002).

\bibitem{Dill90}
K.~A. Dill, Biochemistry {\bf 29},  7133 (1990).

\bibitem{southall_2002}
N.~T. Southall, K.~A. Dill, A.~D.~J. Haymet,  J. Phys. Chem. B {\bf 106}, 521 (2002).

\bibitem{dataset}
The data set is available under request. 

\bibitem{SCOP}
A.~Murzin, S.~Brenner, T.~Hubbard, C.~Chothia, J. Mol. Biol. {\bf 247}, 536 (1995).

\bibitem{Pastor04}
M.~Bogu$\tilde{\mbox{n}}$\'a, R.~Pastor-Satorras, A.~Vespignani, Eur. Phys. J. B {\bf 38},  
  205 (2004).

\bibitem{PKarplus}
P.~A. Karplus,  Protein Sci. {\bf 6}, 1302 (1997).

\bibitem{Silverman03}
R.~Zhou, B.~D. Silverman, A.~K. Royyuru, P.~Athma,  Proteins {\bf 52}, 561 (2003).

\bibitem{alves_2005}
N.~A. Alves, V.~Aleksenko, U.~H.~E. Hansmann,  J. Phys.: Condens. Matter {\bf 17}, 
  S1595 (2005).

\bibitem{Palliser_scales}
C.~C. Palliser, D.~A.~D. Parry, Proteins
  {\bf 42}, 243 (2001).

\bibitem{HZhouYZhou}
H.~Zhou, Y.~Zhou,  Proteins {\bf 49}, 483 (2002).

\bibitem{HZhouYZhou2}
H.~Zhou, Y.~Zhou, Proteins {\bf 54}, 315 (2004).

\bibitem{NewmanPRL02}
M.~E.~J. Newman, Phys. Rev. Lett. {\bf 89},   208701 (2002).

\bibitem{PastorPRL01}
R.~Pastor-Satorras, A.~V\'azquez, A.~Vespignani,  Phys. Rev. Lett. {\bf 87}, 258701 (2001).

\bibitem{PastorPRE03}
M.~Bogu$\tilde{\mbox{n}}$\'a, R.~Pastor-Satorras, Phys. Rev. E {\bf 68}, 036112 (2003).

\bibitem{vasquez_2003}
A.~V\'azquez, M.~Bogu$\tilde{\mbox{n}}$\'a, Y.~Moreno, R.~Pastor-Satorras,   A.~Vespignani, Phys. Rev. E {\bf 67}, 046111 (2003).

\bibitem{fersht_pnas}
A.~R. Fersht, Proc. Natl. Acad. Sci. U.S.A. {\bf 92}, 10869 (1995).

\bibitem{NewmanPRE6703}
M.~E.~J. Newman, Phys. Rev. E {\bf 67}, 026126 (2003).

\bibitem{NewmanStatP10100}
M.~E.~J. Newman,  J. Stat. Phys. {\bf 101}, 819 (2000).

\bibitem{Gromiha_Prog04}
M.~M. Gromiha, S.~Selvaraj, Prog. Biophys. Mol. Biol. {\bf 86}, 235 (2004).

\end{thebibliography}
\end{document}